\documentstyle[12pt,epsf]{article}

\begin{document}

\begin{titlepage}

\font\fortssbx=cmssbx10 scaled \magstep2
\hbox to \hsize{
{\fortssbx University of Wisconsin - Madison}
\hglue.5in$\vcenter{\hbox{\bf MADPH-95-919}
                \hbox{\bf hep-ph/9512248}
                \hbox{December, 1995}}$ }

\vskip 1.in

\begin{center}
{\Large \bf Colorless States in Perturbative QCD: \\
Charmonium and Rapidity Gaps}
\vskip 1.cm
J.\ F.\ Amundson\footnote{Email: amundson@phenom.physics.wisc.edu},
O.\ J.\ P.\ \'Eboli\footnote{Permanent address: Instituto de
F\'{\i}sica, Universidade de S\~ao Paulo,
C.P.\ 66318, CEP 05389-970 S\~ao Paulo, Brazil.
E-mail: eboli@phenom.physics.wisc.edu},
E.\ M.\ Gregores\footnote{Email: gregores@phenos.physics.wisc.edu}, and
F.\ Halzen\footnote{Email: halzen@phenxh.physics.wisc.edu} \\
\medskip
Physics Department, University of Wisconsin, Madison, WI 53706, USA
\end{center}

\vskip 1.in

\begin{center}
{\bf Abstract}
\end{center}

We point out that an unorthodox way to describe the production of
rapidity gaps in deep inelastic scattering, recently proposed by
Buchm\"uller and Hebecker, suggests a description of the production of
heavy quark bound states which is in agreement with data.  The
approach questions the conventional treatment of the color quantum
number in perturbative QCD.

\end{titlepage}

\begin{figure}
\begin{center}
\leavevmode\epsfbox{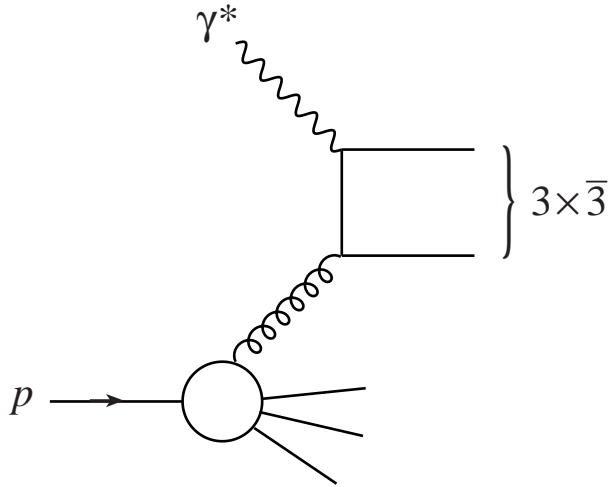}
\end{center}
\caption{\protect\small\baselineskip 12pt
Mechanism for the production of rapidity gaps in
deep inelastic scattering.}
\label{gap:dis}
\end{figure}

Buchm\"uller and Hebecker recently proposed a rather unconventional
description of the formation of rapidity gaps in deep inelastic
scattering \cite{bucheb}.  The mechanism is illustrated in Fig.\
\ref{gap:dis}. The diagram shown represents the production of
final state hadrons which are ordered in rapidity.  From top to bottom
we find the fragments of the intermediate partonic quark-antiquark
state and those of the target. Buchm\"uller and Hebecker proposed that
the origin of a rapidity gap corresponds to the absence of color
between photon and proton, {\em i.e.}  the $\mbox{\bf 3} \times
\bar{\mbox{\bf 3}}$ ($= \mbox{\bf 1} + \mbox{\bf 8}$) intermediate
quark-antiquark state is in a color singlet state.  Because color is
the source of hadrons, only the color octet states yield hadronic
asymptotic states.  This leads to the simple, and experimentally
verified, prediction that
\begin{equation}
F_2^{(gap)} = \frac{1}{1+8} F_2
\end{equation}
This relation embodies the inevitable conclusion that events with and
without gaps are described by the same short-distance dynamics.
Essentially non-perturbative final-state interactions dictate the
appearance of gaps whose frequency is determined by simple counting.

\begin{figure}
\begin{center}
\leavevmode\epsfbox{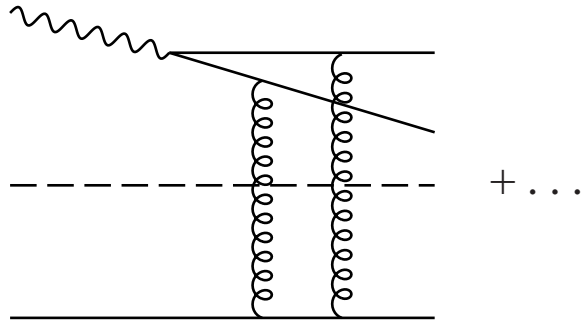}
\end{center}
\caption{\protect\small\baselineskip 12pt
Pomeron mechanism for
the formation of rapidity gaps.}
\label{gap:pom}
\end{figure}

\begin{figure}
\begin{center}
\leavevmode\epsfbox{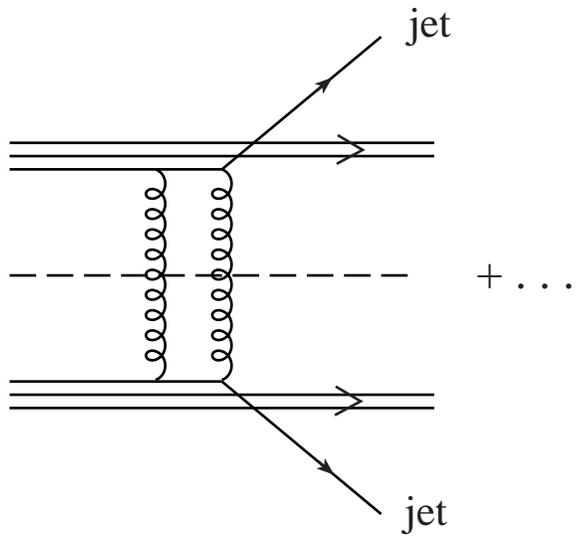}
\end{center}
\caption{\protect\small\baselineskip 12pt
Pomeron mechanism for
the formation of rapidity gaps in hadron collisions.}
\label{2j:pom}
\end{figure}

\begin{figure}
\begin{center}
\leavevmode\epsfbox{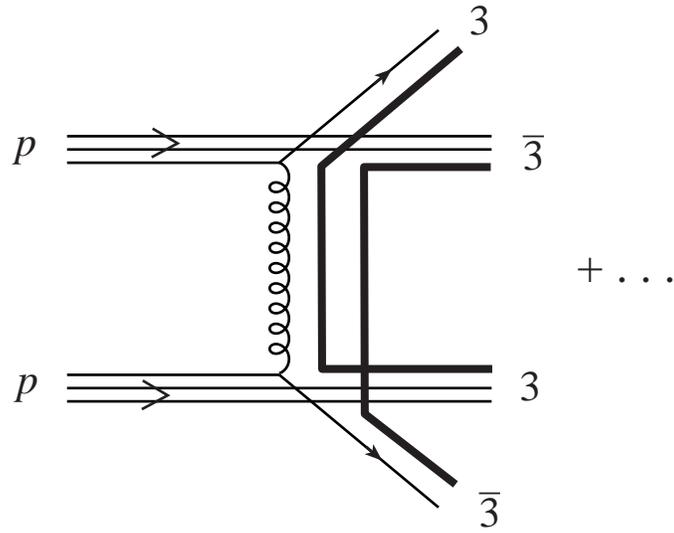}
\end{center}
\caption{\protect\small\baselineskip 12pt
Color bleaching  picture for
the formation of rapidity gaps in hadron collisions.}
\label{2j:ble}
\end{figure}

The orthodox description of rapidity gaps is sketched in Fig.\
\ref{gap:pom}.  The $t$-channel exchange of a pair of gluons in a
color singlet state is the origin of the gap.  The color
string which connects photon and proton in diagrams such as the one in
Fig.\ \ref{gap:dis}, is absent and no hadrons are produced in the
rapidity region separating them.  The same mechanism predicts rapidity
gaps between a pair of jets produced in hadronic collisions; see Fig.\
\ref{2j:pom}.  These have been observed and occur with a frequency of
order of one percent \cite{d0}.  These gaps can, however, be
accommodated as a mere final state color bleaching phenomenon \`{a} la
Buchm\"uller and Hebecker.  This can be visualized using the diagram
shown in Fig.\ \ref{2j:ble}.  At short distances it represents a
conventional perturbative diagram for the production of a pair of
jets.  Also shown is the string picture for the formation of the final
state hadrons.  Color in the final state is bleached by strings
connecting the ${\mbox{\bf 3}}$ jet at the top with the
$\bar{\mbox{\bf 3}}$ spectator di-quark at the bottom and vice-versa.
The probability to form a gap can be counted {\em \`{a} la}
Buchm\"uller and Hebecker to be $1/(1+8)^2$ because it requires the
formation of singlets in 2 strings.  This is consistent with
observation and predicts that, as was the case for electroproduction,
the same short distance dynamics governs events with and without
rapidity gaps.  The data \cite{d0} is consistent with the prediction
of this simple picture and, in fact, hard to understand with the
alternative mechanism of Fig.\ \ref{gap:pom}. The detailed dynamics
may, however, be more complex and, for instance, also involve initial
state radiation.

It is clear from Figs.\ \ref{gap:pom}--\ref{2j:ble} that we have
formulated alternative $s$- and $t$-channel pictures to view the same
physics.  Although they seem at first radically different, this may
not be the case.  Computation of the exchange of a pair of colorless
gluons in the $t$-channel is not straightforward and embodies all the
unsolved mysteries of constructing the ``Pomeron'' in QCD. In a class
of models where the Pomeron is constructed out of gluons with a
dynamically generated mass \cite{natale,chehime}, the diagram of
Fig.\ \ref{2j:pom} is, not surprisingly, dominated by the
configuration where one gluon is hard and the other soft.  The diagram
is identical to the standard perturbative diagram except for the
presence of a soft, long-wavelength gluon whose only role is to bleach
color.  Its dynamical role is minimal, events with gaps are not really
different from events without them.  Soft gluons readjust the color at
large distances and long times.  Their description is outside the
realm of perturbative QCD. Clearly this view of the $t$-channel
exchange is completely compatible with the $s$-channel picture.  At
the perturbative level the $s$- and $t$-channel views are identical
and the color structure of the event is dictated by large distance
interactions: string fragmentation in the $s$-channel and
long-wavelength soft gluons in the $t$-channel picture.  In this class
of models the genuinely hard Pomeron is expected to be no more than an
order $\alpha_s^2$ correction, a view which can be defended on more
solid theoretical ground \cite{cudell}.  Note that our discussion is
at best indirectly relevant to completely non-perturbative phenomena
like elastic scattering.  There is no short distance limit defined by
a large scale.  The Pomeron exists.

        Although we have now made a connection between both pictures,
we have also made a fundamental shift in the way color is viewed in
perturbation theory. It should, in fact, be ignored.  Color structure
is dictated by large-distance fluctuations of quarks and gluons. It is
complex enough that the occupation of different states probably
respects statistical counting, {\em e.g.}, in
determining the fraction of singlet and octet states in a given
rapidity region. In retrospect it does, in fact, seem illogical to
worry about color at short distances, given that soft partons have an
infinite time to readjust any color structure previous to the
formation of asymptotic states.  The main point of this paper is to
suggest that the issue can be settled by studying the production of
colorless states which are theoretically, and experimentally, more
well-defined than rapidity gaps: heavy quark bound states.

\begin{figure}
\begin{center}
\leavevmode\epsfbox{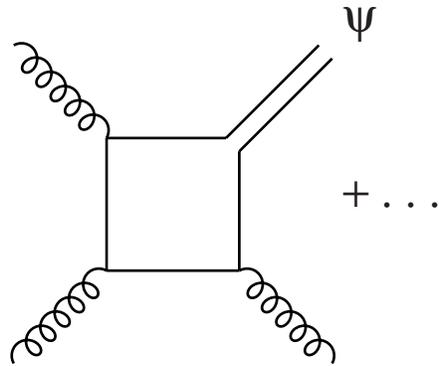}
\end{center}
\caption{\protect\small\baselineskip 12pt
Mechanism for the production
of $J/\psi$ in the color singlet model.}
\label{psi:csm}
\end{figure}

\begin{figure}
\begin{center}
\leavevmode\epsfbox{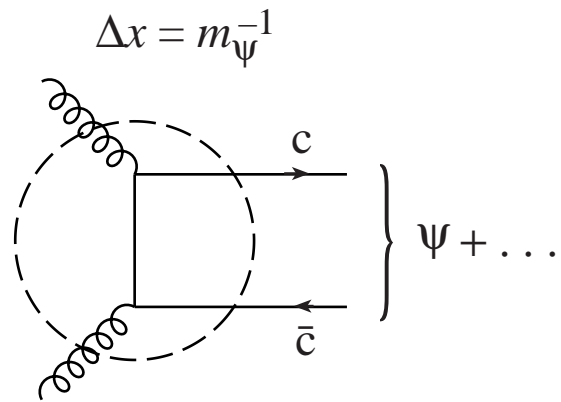}
\end{center}
\caption{\protect\small\baselineskip 12pt
Mechanism for the production
of $J/\psi$ in the color bleaching model.}
\label{psi:cbm}
\end{figure}

In Figs.\ \ref{psi:csm} and \ref{psi:cbm} we show typical diagrams for
the production of $\psi$-particles using the competing treatments of
the color quantum number.  In the diagram of Fig.\ \ref{psi:csm} the
$\psi$ is associatively produced with a final state gluon which is
required to conserve color.  The diagram is related by crossing to the
hadronic decay $\psi \rightarrow 3$ gluons.  In the alternative
approach, however, the color singlet property of the $\psi$ is not
enforced at the perturbative level.  The $\psi$ can, for instance, be
produced to leading order by gluon-gluon, as well as quark-antiquark,
annihilation into $c\bar{c}$.  The latter mechanism is the
color-equivalent of the Drell-Yan process.  Both mechanisms can be
calculated perturbatively.  Their dynamics are dictated by
short-distance interactions of range $\Delta~x \simeq m_{\psi}^{-1}$.
At large distances the exchange of soft gluons between the $c$ and
$\bar{c}$ and the spectators partons will dictate the final color of
the pair.  As before, the probability that a singlet is formed and
produces an onium-state is $1/(1+8)$; the colored states result in the
production of $D \bar{D}$ pairs.  We predict
\begin{eqnarray}
\sigma_{onium} &=& \frac{1}{9} \int_{2 m_c}^{2 m_D} dm~
\frac{d \sigma_{c \bar{c}}}{dm}
\label{sig:on}
\\
\sigma_{open} &=& \frac{8}{9}  \int_{2 m_c}^{2 m_D} dm~
\frac{d \sigma_{c \bar{c}}}{dm}
+ \int_{2 m_D} dm~\frac{d \sigma_{c \bar{c}}}{dm}
\label{sig:op}
\\
& \simeq &\frac{8}{9}  \int_{2 m_c}^{2 m_D} dm~
\frac{d \sigma_{c \bar{c}}}{dm}
\nonumber
\label{sig:ap}
\end{eqnarray}
where $\sigma_{c\bar{c}}$ is the cross section for producing heavy
quark pairs.  It is computed perturbatively.  These relations can be
generalized in an obvious way to differential cross sections, such as
$d\sigma/dx_F$ and $d\sigma/dp_T$.

The approach embodied in Eqs.\ (\ref{sig:on}) and (\ref{sig:op}) seems
reasonable: it is indeed hard to imagine that a color singlet state
formed at a range $m_{\psi}^{-1}$, as in Fig.\ \ref{psi:csm}, automatically
survives to form a $\psi$.  This formalism was, in fact, proposed
almost twenty years ago \cite{cem,fh:1,gor} and subsequently abandoned
for no good reason.  Maybe the time has come to contemplate the
possibility that the phenomenological problems \cite{braaten} of the
current perturbative approach in describing data on the hadronic
production of onium-states is connected to the inappropriate treatment
of color.  Other approaches of similar spirit can be found in Refs.\
\cite{bbl} and \cite{hoyer}.

This also raises the question whether the alternative approach is
(still) consistent with data.  In the limit that the masses $m_\eta$,
$m_\psi$ and $2 m_D$ are equal, Eqs.\ (2) and (3) state that the
production of onium- and open charm states is dictated by identical
dynamics.  Only a normalization factor connected to color
differentiates the two cases.  This way of viewing the approach
mirrors Eq.\ (1) for deep inelastic scattering.  One may be able to
take the idea of final state counting one step further by dividing the
total color-singlet cross section, which is rigorously predicted, into
onium-states according to the simple statistical counting
\begin{equation}
\sigma_X = \rho_X \sigma_{onium}
\label{frac}
\end{equation}
with
\begin{equation}
\rho_X = \frac{2 J_X + 1}{\sum_i (2 J_i +1) } \; ,
\label{spin}
\end{equation}
where $J_{X}$ is the spin of any onium state $X$; the sum runs over
all onium states.

The above formalism can, at best, be approximate; we expect phase space
corrections favoring the lighter states. It is, in this context,
interesting to note that any approach must satisfy the sum rule that the
sum of the cross sections of all onium-states is given by Eq.\ (2),
{\em i.e.}
\begin{equation}
\sum_i \sigma_i = \frac{1}{9} \int_{2 m_c}^{2 m_D} dm~
\frac{d \sigma_{c \bar{c}}}{dm} \; .
\end{equation}
This relation is, unfortunately, difficult to test experimentally
since it requires measuring cross sections for {\em all} of the
charmonium bound states at a given energy.

\begin{figure}
\begin{center}
\leavevmode\epsfbox{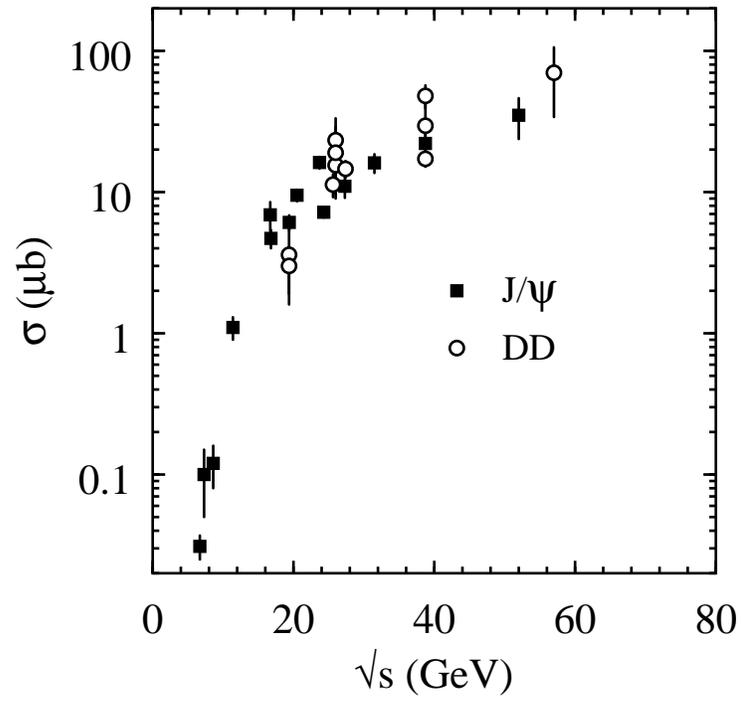}
\caption{ \protect\small\baselineskip 12pt
Total cross section per nucleon for the
production of $D\bar{D}$ (circles) and $J/\psi$ (squares) in
proton-nucleon collisions. The $J/\psi$ cross section has been multiplied
by a constant factor of $50$.}
\label{fig:dyn}
\end{center}
\end{figure}

\begin{figure}
\begin{center}
\leavevmode\epsfbox{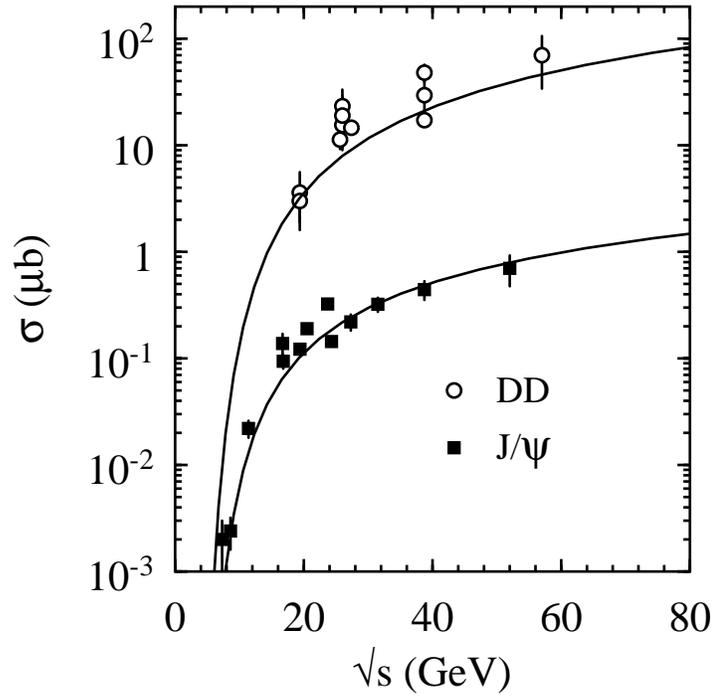}
\caption{ \protect\small\baselineskip 12pt
Total cross section per nucleon for the
production of $D\bar{D}$ (circles) and $J/\psi$ (squares) in
proton-nucleon collisions.}
\label{sig:th}
\end{center}
\end{figure}

\begin{figure}
\begin{center}
\begin{tabular}{cc}
        \leavevmode\epsfbox{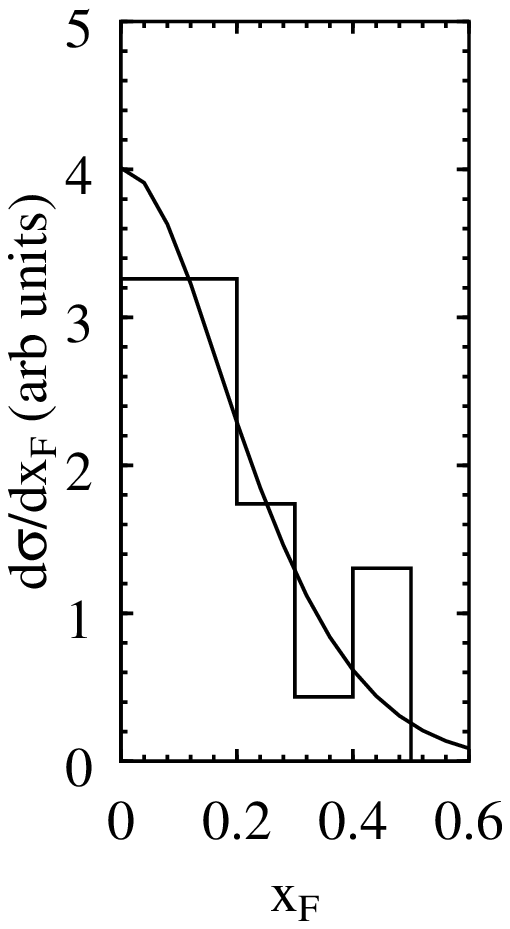} & \leavevmode\epsfbox{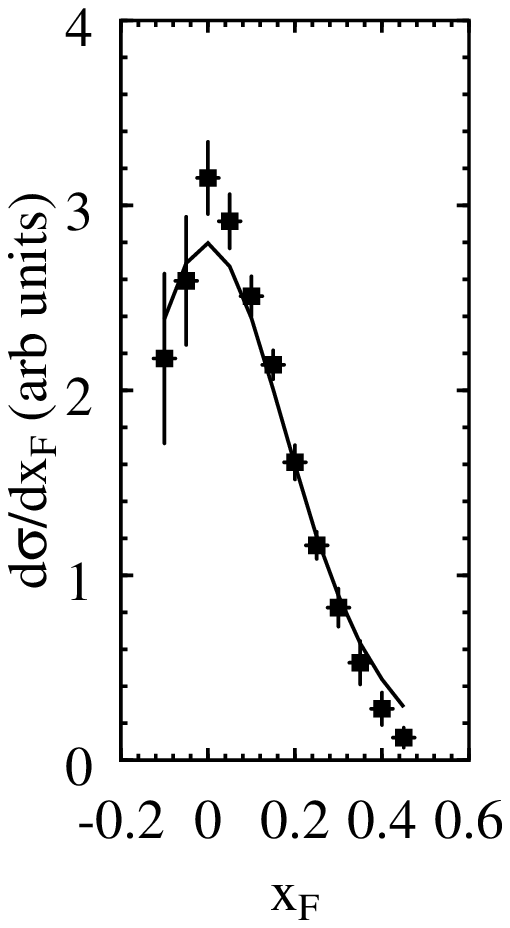} \\
\end{tabular}
\caption{ \protect\small\baselineskip 12pt Normalized $x_F$
distribution for the production of of $D\bar{D}$ (histogram) at
$\protect\sqrt{s} = 27.4$ GeV and $J/\psi$ (squares) in proton-nucleon
collisions at $\protect\sqrt{s} = 23.7$ GeV. The curves are the
prediction of our model.}
\label{fig:xf}
\end{center}
\end{figure}

Our prediction of Eqs.\ (\ref{sig:on}) and (\ref{sig:op}) that the
production of hidden and open charm have similar dynamics is supported
by Fig.\ \ref{fig:dyn}, where we have plotted the experimental data
for the production cross section of $J/\psi$ \cite{exp:psi} and
$D\bar{D}$ \cite{exp:charm} as a function of the center-of-mass
energy.  The $J/\psi$ data has been multiplied by a constant, which,
in the aforementioned limit where $m_{\psi} = 2m_{D}$, is simply
$8/\rho_{\psi}$.  The agreement is remarkable.  Obviously, higher
statistics data would exhibit deviations from our predictions since
threshold effects have not been taken into account.  In Fig.\
\ref{sig:th} we directly compare our leading-order predictions of
Eqs.\ (\ref{sig:on}) and (\ref{frac}) with data on $J/\psi$ and
$D\bar{D}$ production.  We here used the MRS A parameterization of the
nucleon structure function with the scale $Q^2 = \sqrt{\hat{s}}$ and
$m_c = m_\eta/2$ in our calculations.  In order to fit the data, we
multiplied the leading order prediction of Eq.\ (\ref{sig:op}) by a
``K'' factor of $K_{D\bar D}=6.4$.  We multiplied the prediction of Eq.\
(\ref{sig:on}) by a factor of $\rho_{\psi}K_{onium} = 3.0$.  It is
interesting to notice that the energy dependencies of the cross
sections are well explained by our simple model.  Assuming that the
next to leading order corrections ($K$ factors) to both processes are
equal, {\em i.e.} $K_{D\bar D}=K_{onium}$, we determine the fraction of
$J/\psi$ ($\rho_\psi$): $\rho_\psi = 3.0/6.4 = 0.47$.  In the same
spirit, starting from Eqs.\ (\ref{sig:on}) -- (\ref{sig:ap}), our model
predicts that the normalized $x_F$ distribution for $J/\psi$ and
$D\bar{D}$ pairs should be the same at a given center-of-mass energy.
This is indeed the case \cite{exp:xfddbar,exp:xfpsi}; see Fig.\
\ref{fig:xf}.

In our model the production of charmonium states proceeds in two
stages.  First the $c\bar{c}$ pair is produced perturbatively,
subsequently it evolves non-perturbatively into the asymptotic states.
Therefore, the study of the production of specific onium-states should
shed light on the non-perturbative dynamics responsible for onium
formation.  The available experimental data \cite{exp:frac} for the
ratios $\sigma(\chi) / \sigma(\psi)$ and $\sigma(\psi^\prime) /
\sigma(\psi)$ indicate that these quantities are not only independent
of the center-of-mass energy but also independent of the target and
incident beams \cite{gavai}.  We summarize the experimental results
for fractions of individual charmonium states in Table \ref{tab:frac}.
Simple counting predicts that the fraction of $c\bar{c}$ pair forming
a given state is given by (\ref{spin}).  However, this does not take
into account the feed-down from the decays of heavier states into the
lighter ones.  In Table \ref{tab:frac} we also list the predictions
corrected for this effect.  This very simple model clearly captures the
general features of the data.  A quantitative comparison requires a
model which incorporates phase space.  It is nevertheless important to
point out that simple counting predicts a substantial production of
$\psi^\prime$ in agreement with observations.  This should be
contrasted with the conventional treatment.

A more detailed discussion of both theory and experiment, including
applications to bottomonium, will be published elsewhere.

We would like to thank C.\ O.\ Escobar and I.\ Bediaga for helping
locate experimental results on charm.  This research was supported in
part by the University of Wisconsin Research Committee with funds
granted by the Wisconsin Alumni Research Foundation, by the U.S.\
Department of Energy under grant DE-FG02-95ER40896, and by Conselho
Nacional de Desenvolvimento Cient\'{\i}fico e Tecnol\'ogico (CNPq).



\begin{table}
\begin{center}
\begin{tabular} {|c|c|c|}
\hline
Observable &Experimental results &Statistical model \\
\hline
Fraction of $\psi$ from $\chi$   decay      & $~0.36\pm0.02~$ & 0.24 \\
Fraction of $\psi$ from $\psi^\prime$ decay & $0.059\pm0.005$ & 0.28 \\
$\sigma(\psi^\prime)/\sigma(\psi) $         & $~0.10\pm0.01~$ & 0.48 \\
$\sigma_{direct}(\psi)$                     & $~0.57\pm0.02~$ & 0.48 \\
\hline
\end{tabular}
\caption{\protect\small\baselineskip 12pt Components of the charmonium
production.  We have averaged the available experimental data.}
\label{tab:frac}
\end{center}
\end{table}


\protect


\protect


\protect


\protect


\protect


\protect


\end{document}